\documentclass{ieeeaccess}
\usepackage{cite}
\usepackage{amsmath,amssymb,amsfonts}
\usepackage{algorithmic}
\usepackage{graphicx}
\usepackage{textcomp}
\def\BibTeX{{\rm B\kern-.05em{\sc i\kern-.025em b}\kern-.08em
    T\kern-.1667em\lower.7ex\hbox{E}\kern-.125emX}}
\begin{document}
\history{Date of publication xxxx 00, 0000, date of current version xxxx 00, 0000.}
\doi{10.1109/ACCESS.2017.DOI}

\title{Federated 3GPP Mobile Edge Computing Systems: A Transparent Proxy for Third Party Authentication with Application Mobility Support}
\author{\uppercase{Asad Ali}\authorrefmark{1}, \IEEEmembership{Student Member, IEEE},
\uppercase{Samin Rahman Khan}\authorrefmark{2},
\uppercase{Sadman Sakib}\authorrefmark{2},
\uppercase{Md. Shohrab Hossain}\authorrefmark{2}, \IEEEmembership{Member, IEEE},
and \uppercase{Ying-Dar Lin}\authorrefmark{1} \IEEEmembership{Fellow, IEEE}.}
\address[1]{Department of Computer Science, National Yang Ming Chiao Tung University, Hsinchu, Taiwan (e-mail: ali.eed06g@nctu.edu.tw; ydlin@cs.nctu.edu.tw)}
\address[2]{Department of Computer Science and Engineering, Bangladesh University of Engineering and Technology, Bangladesh (e-mail: saminrahmankhan97@gmail.com; saadsakib3@gmail.com; mshohrabhossain@cse.buet.ac.bd)}

\tfootnote{This work was supported by the Ministry of Science and Technology, Taiwan under Project 108-2221-E-009-046-MY2.}

\markboth
{Author \headeretal: Preparation of Papers for IEEE TRANSACTIONS and JOURNALS}
{Author \headeretal: Preparation of Papers for IEEE TRANSACTIONS and JOURNALS}

\corresp{Corresponding author: Samin Rahman Khan (e-mail: saminrahmankhan97@gmail.com).}

\begin{abstract}
Multi-Access or Mobile Edge Computing (MEC) is being deployed by 4G/5G operators to provide computational services at lower latencies. Federating MECs across operators expands capability, capacity, and coverage but gives rise to two issues\textemdash third-party authentication and application mobility\textemdash for continuous service during roaming without re-authentication. In this work, we propose a Federated State transfer and 3rd-party Authentication (FS3A) mechanism that uses a transparent proxy to transfer the information of both authentication and application state across operators to resolve these issues. The FS3A proxy is kept transparent, with virtual counterparts, to avoid any changes to the existing MEC and cellular architectures. FS3A provides users with a token, when authenticated by an MEC, which can be reused across operators for faster authentication. Prefetching of subscription and state is also proposed to further reduce the authentication and application mobility latencies. We evaluated FS3A on an OpenAirInterface (OAI)- based testbed and the results show that token reuse and subscription prefetching reduce the authentication latency by 53\textendash65\%, compared to complete re-authentication, while state prefetching reduces application mobility latency by 51\textendash91\%, compared to no prefetching. Overall, FS3A reduces the service interruption time by 33\%, compared to no token reuse and prefetching.
\end{abstract}


\titlepgskip=-15pt

\maketitle

\section{Introduction}
\label{sec:introduction}
\PARstart{E}{dge} Computing is the computing paradigm where computation and data storage services are brought to the edge of the network, closer to its users \cite{shi2016edge}. Multi-Access or Mobile Edge Computing (MEC) is a concept that integrates computational capability with cellular networks and brings enhanced computational power closer to the users \cite{siriwardhana2021survey}, in order to reduce latency and bandwidth requirements and increase the quality of service. Real time applications like augmented reality, online gaming, video conferencing, video processing for autonomous vehicles etc., benefit considerably from low latency edge systems.

As the Internet of Things (IoT) continues to grow, a large number of mobile IoT devices will turn to Edge Computing and Mobile Edge computing for computation capability and networking \cite{liu2020toward}. Third Generation Partnership Project (3GPP) networks like 4G Long Term Evolution (LTE) or 5G provide the connectivity backbone for MEC. Mobile Network Operators (MNOs) can easily incorporate MEC into their existing infrastructure and provide services to their users \cite{mece5g2020}. The European Telecommunications Standards Institute (ETSI) has provided many new use cases for MEC and has pointed out that the deployment of MEC will introduce new revenue streams for operators, vendors and third-parties \cite{mecp2018}. 

\subsection{Federation}
There are multiple MNOs in the world and they likely have their own MEC servers deployed in their infrastructure. The subscribers of these MNOs would be able to access the services provided by the deployed MEC servers. In the early deployment stage, individual MNOs may not be able to provide MEC coverage in all areas. Different MNOs would cover different areas, depending upon the location of MEC servers. Therefore, a user will not be able to get complete coverage and, if there is no collaboration among multiple MNOs, a subscriber would have to buy a subscription from multiple MNOs and create multiple accounts for multiple application servers deployed in different MNOs. Furthermore, having to provide login credentials every time a user tries to access an application would greatly increase the latency and reduce a user's experience.

Hence, it would be more advantageous for both the service providers and their subscribers, if the service providers could form a federation among themselves. This kind of federation is already in practice for cloud service providers \cite{usama2021agg}. Such a federation would encourage MNOs to collaborate and share resources among themselves in order to provide multiple services and better coverage to all their users. Such a federation among MNOs is better as compared to a federation among MNOs and cloud or fog networks because the base stations of MNOs are closer to each other. Therefore, a federation among MNOs would allow them to share resources and provide faster mobility to their users. Furthermore, a federation among MNOs would provide extended MEC service coverage.

The advantages of such a federation are twofold: MNOs would be able to serve more users and users will have access to continuous low latency MEC services without having to rely on cloud or fog. Users would be able to obtain MEC services from such a federation using just a single SIM card without having to buy subscriptions from different MNOs, and they would also be able to switch between networks in order to get the best services. A federated system would also open  doors to more modularised services and applications that can be shared by multiple networks.

\subsection{Issues and Research Questions}
\subsubsection{Authentication Issue}
User equipment (UE) can authenticate itself with an MEC via its cellular authentication information which is stored in the cellular network it is registered with, which we refer to as the home network of the UE. Whenever the UE wants to use MEC services from any other network (referred as the foreign network), the MEC system needs to access the authentication information from the UE’s home network in order to verify its identity. The issue that arises here is that the authentication information of the UE is kept private in the home network and is not shared outside the network’s trust domain;  providing interfaces to expose the authentication information outside the cellular network would demand changes in the already established cellular network architecture and protocols. This gives rise to the third-party authentication issue where we need a mechanism by means of which the UE (first party) could authenticate itself with the foreign MEC (third party) via its home MEC (second party).

\subsubsection{Application Mobility Issue}
When an UE moves out of the MEC coverage of the home network, and gets authenticated with the foreign network, we are faced with another issue we refer to as the application mobility issue. The issue is that the UE  was getting services from the MEC of home network and now it has moved to the foreign network where it needs to get the services from the MEC platform of the foreign network without discontinuation. If a user cannot move instantaneously into the MEC coverage of neighboring edge networks, a discontinuation of services would occur, causing the UE to fall back to the cloud servers which would, in turn, add latency and degrade the user’s experience. Moreover, whenever a user switches edge networks, active application sessions must be retained because the user’s convenience will be greatly degraded if the user has to login again in a new network and start the session anew. The application state should be kept intact and latency during transfer should also be kept to a minimum. Therefore, an MEC platform in one network needs to be able to access user application state from other MEC platform in another network. This is the application mobility issue and we need a mechanism to transfer the user session state from one MEC to another MEC located within different networks.

\subsubsection{Research Questions}
In summary, we have identified two main issues for the federated 3GPP MEC systems: 3rd party authentication and application mobility. In this work, our goal is to find the answer to the following question: how do we support low latency MEC application authentication and mobility in a federation among multiple 3GPP MEC systems while maintaining transparency? In order to solve the identified  issue, we need an intermediary body that can facilitate internetwork MEC communication, third-party authentication, and state transfer. All this has to be done while maintaining transparency and conforming to the guidelines provided by the ETSI. To come up with a transparent solution, we first look at 3GPP and ETSI standards. ETSI has provided some design specifications for MEC architecture \cite{mect2019}, security, authentication, application mobility \cite{mece2e2017}\cite{meceam2020}, and others. We consider these design specifications while proposing the solution to the identified issues.

\subsection{Transparent Proxy Solution And Latency Reduction Optimizations}
We have explored literature that addresses authentication and application mobility issues in federated mobile networks. To the best of our knowledge, no study in the current literature provides solutions to both the issues identified in the previous section. There are two different ways to solve the two issues we have identified; one is to propose a new protocol that provides authentication and application mobility across multiple MNOs and the other way is to use the existing 3GPP standard protocols and build the solution using the existing protocols. The latter is better as it does not involve any modifications in the existing protocols and provides transparency.

Solutions have been proposed for developing a transparent, low latency system that provides Authentication, Authorization, and Accounting (AAA) functionality in a MEC system in a single network \cite{li2020transparent}. Transparent solutions have also been proposed for third-party authentication and application mobility between MEC platforms in a single network \cite{ali2020transparent}. Our proposal integrates these solutions and provides solutions to authentication and application mobility problems in federated 3GPP inter-MNO MECs, is transparent, and meets the low latency requirements in a federated 3GPP MEC System.

We propose a Federated State transfer and 3rd-party Authentication (FS3A) mechanism that makes use of a transparent proxy that can transfer the authentication and application state information across the MNOs to provide 3rd-party authentication and application mobility. The basic design idea is based on a proxy network that keeps inter-MNO communications close to the ground and leverages roaming based cellular authentication to provide 3rd party application authentication in foreign networks. The proxy is consists of virtual counterparts, so as to avoid any changes to the existing MEC and 4G/5G cellular network architectures. UEs are authentication by a foreign MEC network by inspecting cellular traffic during roaming authentication in the foreign cellular network. The MEC network can then provide access tokens to the UEs for accessing its application servers. FS3A also provides scope for reusing application access tokens across the MECs that belong to other MNOs for faster authentication. In FS3A,  MEC-tier application state transfer is done using MEC host and system-level entities to manage the application mobility in the federation. Subscription data and application state data prefetching are done to further reduce the authentication and application mobility latencies. The main contributions of this work are summarized as follows:

\begin{itemize}
    \item We propose a federation between MECs deployed by different MNOs and develop an FS3A mechanism for solving the third-party authentication and application mobility issues in the proposed federation.
    \item The FS3A makes use of a proxy network, host and system level MEC entities to transfer the authentication and application state information across MECs in different MNOs. These adaptations are compliant with current MEC specifications, existing 4G LTE network standards and will be compatible with 5G standards as well.
    \item FS3A demonstrates how low-latency application authentication is provided in foreign MEC systems by inspecting roaming cellular authentication. FS3A provides scope for reusing application access token without re-authentication through MEC system, reducing application down-time while moving across networks.
    \item FS3A performs subscription data and application state data prefetching to further reduce the authentication and application mobility latency.
\end{itemize}

The rest of the paper is organised as follows. In Section 2, we review the 3GPP and ETSI standards, look into the threat model for our problem and explore relevant literature to learn how similar problems were solved. In Section 3, we state the problem and provide some examples to illustrate various scenarios, and in Section 4 we described in detail our approach, architecture design, and message flows. Section 5 covers the details of the implementation, modules, and testbed. Section 6 reflects the results and evaluation them. We conclude the paper in Section 7 and present some points to be investigated in the future.

\section{Background and Related Work}
In this section we discuss existing 4G LTE architecture with MEC, some of the relevant ETSI MEC standards, a threat model for our problem, and related work.

\subsection{Existing 4G-LTE Architecture with MEC}

4G-LTE is the most widespread mobile network in the world and is a prime candidate for an underlying cellular network within which MEC systems will function. Designs developed for an LTE network will also have to be compatible with future 5G networks. An UE, eNodeB (eNB), and an Evolved Packet Core (EPC) constitute an LTE system. EPC consists of a Home Subscriber Server (HSS), Mobility Management Entity (MME), and Serving and Packet Gateway (SPGW). An UE is the end-device through which a user communicates with an EPC via eNB. The HSS acts as the database of all cellular information of subscribers. The MME manages the access and mobility aspects of each UE. MEC servers can be deployed in 4G LTE network through different architectures provided by ETSI \cite{giust2018mec}. In order to achieve lower latency, we used the Bump-in-the-wire approach where the MEC platforms are deployed in between the eNBs and the EPC.

\subsection{MEC Entities and ETSI Standards for MEC}
An MEC system consists of different host- and system-level entities which manage the functionalities within a single MEC platform, and the functionalities across all the platforms in the system. The MEC manager is a host level entity that manages the components inside the MEC platform. An MEC Orchestrator (MEO) is a system-level entity that maintains an overview of the MEC system based on deployed mobile edge hosts, available resources, available mobile edge services, and the topology. An MEO also triggers application instantiation, termination, and application relocation at different MEC platforms in order to provide the best services based on the available resources \cite{meceam2020}. As per ETSI standards, application mobility services are to be provided by an MEC systems. MEC systems need to provide necessary information to the MEC applications about when and where to transfer user context during application mobility \cite{etsi2019multi}. This overview of the MEC system explains the functionality of different components and helps to identify how the functionality of these components can be extended to accommodate federated authentication and mobility.

\begin{table*}[h]
\renewcommand{\arraystretch}{1.3}
\caption{Related Work}
\label{table_one}
\centering
\begin{tabular}{|c||c||c||c||c|}
\hline
Author & Method & 3p/Multi-Network Authentication & Application Mobility & Transparency\\
\hline
\hline
Donald\cite{donald2015secure} & Mobile Cloud Authentication & \checkmark & X & \checkmark\\
\hline
Bonnah\cite{bonnah2020decchain}  & Decentralised Blockchain Infrastructure & \checkmark & X & X\\
\hline
Targali\cite{targali2013seamless} & Mutual Key Exchange & \checkmark & X & X\\
\hline
Choyi\cite{choyi2016seamless} & Authentication Proxy & \checkmark & X & X\\
\hline
Edris\cite{edris2020network} & Federated ID, Oauth2 & \checkmark & X & X\\
\hline
Cui\cite{cui2021edge} & Reinforcement Learning & \checkmark & X & \checkmark\\
\hline
Han\cite{han2019handover} & Handover Authentication & \checkmark & X & \checkmark\\
\hline
Danial\cite{shah2020sdn} & SDN & \checkmark & X & X \\
\hline
Mwangama\cite{mwangama2015towards} & SDN & X & \checkmark & X \\
\hline
Pencheva\cite{pencheva2018open} & Application Driven Handover & X & \checkmark & \checkmark\\
\hline
Ours & Transparent Proxy & \checkmark & \checkmark & \checkmark\\
\hline
\end{tabular}
\end{table*}

\begin{table}[!t]
\renewcommand{\arraystretch}{1.3}
\caption{TC3A, TS3A, and FS3A Comparison}
\label{table_two}
\centering
\begin{tabular}{|c||c||c|c|}
\hline
Parameters & TC3A & TS3A & FS3A\\
\hline
3-p Authentication & \checkmark & \checkmark & \checkmark\\
\hline
Application Mobility & \checkmark & \checkmark & \checkmark\\
\hline
State Prefetching & X & \checkmark & \checkmark\\
\hline
Subscription data prefetching & X & X & \checkmark\\
\hline
Inter-MNO Connectivity & X & X & \checkmark\\
\hline
\end{tabular}
\end{table}

\subsection{Threat Model}
In this work, a cellular network and an MEC system are considered as benign. However,  UEs are considered to be malicious. A malicious UE may or may not be a user of the underlying cellular network. Malicious traffic sent by UEs can be handled by the core of an LTE network, but since MEC traffic does not go through the core network, malicious UEs can pose different threats to MEC systems. A malicious user may steal user IDs from other users or may reuse an ID that has already been used by itself or by others to get unauthorized access to MECs. Users may also hijack session IDs of others and may generate large amounts of traffic to incapacitate an MEC system. A federated MEC system should be resilient against these attacks. In section IV, we will see how our design copes against these problems.

\subsection{Related Work}
In order to find solutions to the proposed problem, we examined studies that have proposed solutions to at least one of these problems, either 3rd party authentication in federated systems, multi-network authentications, or application mobility across networks. We analyzed the transparency of these proposals according to existing LTE protocols and MEC standards. Our findings  are summarized in  Table \ref{table_one}. Some studies \cite{donald2015secure} propose third-party authentication through the cloud. There are also studies \cite{bonnah2020decchain}-\cite{cui2021edge} that use blockchain, mutual key exchanges, Reinforcement learning (RL), and federated IDs to manage Multi-network authentication. A study \cite{han2019handover} provides handover authentication by improving the existing EAP-AKA protocol and another study \cite{Niew2021token} makes use of tokens to provide service access control in 5G MEC. A few studies \cite{shah2020sdn}\cite{mwangama2015towards} also use SDNs to manage authentication and application mobility in federated networks, while another \cite{pencheva2018open} provides application driven handover while retaining transparency.

We also looked into  studies that propose solutions for application relocation and most of them focus on VM migration, for example \cite{ouyang2018follow} and \cite{wang2019dynamic}. On the other hand, recent literature also provides transparent, low-latency solutions for authentication and mobility within a single edge network \cite{ali2020transparent}\cite{machen2017live}\cite{aghdai2019enabling}, anonymous mutual authentication in MEC platforms \cite{lee2021anon}, and third-party authentication between cloud and edge \cite{lin2020proxy}. Some work has also been done on the deployment of a secure MEC system in a cellular network that can itself provide authentication for its applications \cite{li2020transparent}. We use this MEC deployment method in our work and extend it further across federated MNOs.

An earlier study from some of us \cite{ali2020transparent} provides low-latency authentication and state transfer while moving from one MEC platform to another by using two approaches, termed TC3A and TS3A. In order to  compare this work with \cite{ali2020transparent}, we summarize the differences in Table \ref{table_two}. It can be seen that TC3A, TS3A, and our proposed FS3A provide third-party authentication and application mobility but FS3A does that in a federated MEC system. The major advantage of FS3A over TS3A and TC3A is that it solves the challenges of subscription and state data prefetching and provides the Inter-MNO MEC connectivity, while TS3A and TC3A work for Intra-MNO MECs that are deployed by the same MNO.

Our survey shows that previous studies have provided solution to authentication and application mobility issues independently in the MECs deployed by an MNO but, no previous study provides solutions for authentication and application mobility issues together while maintaining transparency and latency requirements in federated MEC systems across multiple MNOs. It should be noted that we do not try to solve the call and data roaming problems as they have already been addressed by various studies \cite{roos2003roaming} \cite{inaba2014lte} \cite{mafa20215g}. We solely focus on solving the problems of third party authentication and application mobility for a user that roams between MECs deployed by different MNOs. To the best of our knowledge, ours is the first work that provides a complete solution to the problems at hand, i.e., third-party authentication and application mobility in federated MEC systems while meeting transparency and latency requirements. We adopt the transparent, low-latency authentication solutions from Intra-MNO MEC \cite{li2020transparent}\cite{ali2020transparent} and enhance them to solve the problem of third-party authentication and state transfer in an Inter-MNO MEC federation.

\section{Problem Statement}
Low latency third-party authentication and application mobility in a federated MEC system are the two problems we aim to solve. In this section, we elaborate these problems and present the assumptions we made, and issues related to these problems.

\subsection{Problem Scenarios}
\begin{figure}[!t]
\centering
\includegraphics[width=3.4in]{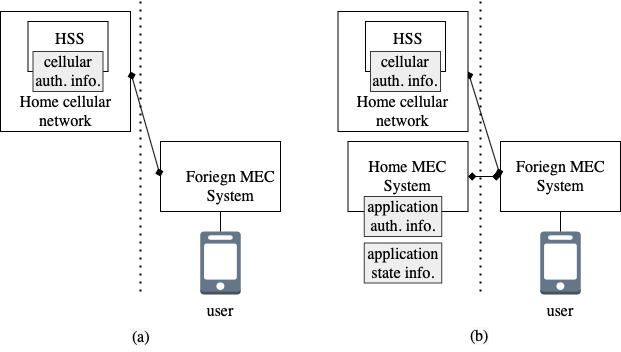}
\caption{(a) Authentication problem for new application; foreign MEC system needs to authenticate UE by using cellular authentication information from the home network (b) Authentication and application mobility problems for application continuation; foreign MEC system needs to get application state from the home MEC system as well.}
\label{fig_description}
\end{figure}

Consider two 4G-LTE MNOs in a federation which provide MEC services. We assume that they use an MECSec design \cite{li2020transparent} to manage authentication, authorization, and access control. We assume too that an UE is the subscriber of one of the MNOs (referred as the home network). The HSS in the home network contains the subscription, authentication, and other information about the UE. The other networks where the UE does not have a subscription are termed foreign networks. In this work, we consider two scenarios.  In the first scenario, we assume that the UE wants to access the MEC services provided by a foreign network, and where the UE needs to obtain authenticated in the foreign MEC with the authentication credentials from the home network, as shown in Fig. \ref{fig_description}(a). 

In the second scenario, as shown in Fig. \ref{fig_description}(b), we assume that the UE has moved to the foreign network while using an MEC service in the home network. In this case, the foreign network derives authentication information either from the home cellular network or from the home MEC system. Application state information also needs to be carried from the application instance in the home network to its foreign counterpart. In this work, we will only consider the scenario where the UE moves from the home to the foreign network. Cases where the user moves from the foreign to the home network have not been considered here, because the UE is already authenticated with the home network and application mobility in this case works the same as the home to foreign networks scenario. We have assumed that the underlying LTE networks are connected for providing roaming cellular services to their users.

\subsection{Low latency and Transparency Challenges}
Some challenges arise when an UE moves in a federated MEC system as it is important for the UE to be identified by the MEC systems in all networks by means of a common ID. The ID for each user must be unique and unchangeable. A UE is then given a unique temporary ID by the cellular network. The entities in the cellular networks also use different temporary IDs to keep track of the tunnels formed to serve the user. The UE is also given an IP address after it has attached to the cellular network. Since the MEC platform sits between the EPC and eNBs, it can only inspect the packets in the S1 interface. An UE may come into an MEC platform either by attaching for the first time by service resumption, or via handover. In these cases the UE attaches to the cellular networks using different IDs. Under all these circumstances, it must be made sure that the traffic from a user is uniquely identified by the MEC system.

The application mobility in MEC systems belonging to different MNOs also creates some challenges. The MEC system must have the following information before it does application state transfer: from where the application state is to be transferred, to whom is it to be transferred, and when to transfer. Since the user becomes completely detached from the home network and attaches to the foreign network, the home network has no way of knowing beforehand to which foreign network the user will move,  and so the MEC platform in the home MEC system cannot initiate the application handover. It is therefore important to solve this state transfer challenge with the lowest possible latency. Another major challenge is to keep the system design transparent so that no modification of the current cellular and MEC infrastructure is required. \emph{In summary, our objective is to provide transparent 3rd-party authentication for the UEs in the foreign networks along with seamless application mobility with minimal latency.}

\section{Proposed Solution And Architecture Design}
In order to solve the third-party authentication and application mobility problems identified in the previous section, along with the low latency and transparency issues, we propose a Federated State transfer and 3rd-party Authentication (FS3A) mechanism. FS3A makes use of a transparent proxy in order to transfer the authentication and application state information of an UE across MECs in different MNOs. The main design idea for the proxy is transparency, and the reason behind the transparent proxy is to avoid any changes in the existing infrastructure of the underlying MEC and cellular architectures. In order to provide transparency, we propose virtual counterparts in the proxy so that the entities from the cellular and MEC systems communicate with their virtual counterparts in the proxy. The FS3A also provides the UEs with a token when they become authenticated by an MEC, which is reused across the MECs deployed by different operators for faster authentication. FS3A also makes use of subscription and state data prefetching to further reduce the authentication and application mobility latencies. The proposed proxy based solution is unique as it not only provides the solutions for third-party authentication and application mobility issues in federated 3GPP inter-MNO MECs but also provides (subscription and state) data prefetching and token reuse to reduce the latency. FS3A is explained in detail in the next subsection.

\subsection{FS3A}
The FS3A has two parts, one is for third-party authentication and other is for application mobility via state transfer. For authentication in the foreign network, the MEC system picks up UE authentication information from the underlying cellular network. We propose adding a system-wide datastore for each individual MEC system for storing all authentication information and user information, making it possible for the MEC system to store and share user information among MEC host platforms during its lifetime in the network. For application mobility via state transfer, we  included two mobility functions, the host level Application Mobility Service (AMS) and the system level Application Mobility Coordinator (AMC) in the MEC system. AMCs exchange user context and state transfer information through the proxy that provides the means for inter-MNO communication. FS3A ensures that no changes are required in the existing protocols and design of the underlying cellular network.

FS3A places most of the responsibility of application state transfer on the mobility management functions (AMS and AMC) which reduces the burden on the MEC applications and UEs. FS3A does not require any assistance or intervention from the user, making it highly secure and fast as all the processes are run in the MEC tier. The proposed FS3A solution also makes use of few optimizations such as the token reuse instead of re-authentication at the foreign network and data prefetching in order to further reduce the latency. Overall, FS3A is transparent in design, fast, and secure from the attacks of  malicious users, and it is also easy to deploy. In the next subsection, we look at the architecture design and individual components that are required for FS3A in more detail.

\subsection{Architecture}
\begin{figure}[!t]
\centering
\includegraphics[width=3.3in]{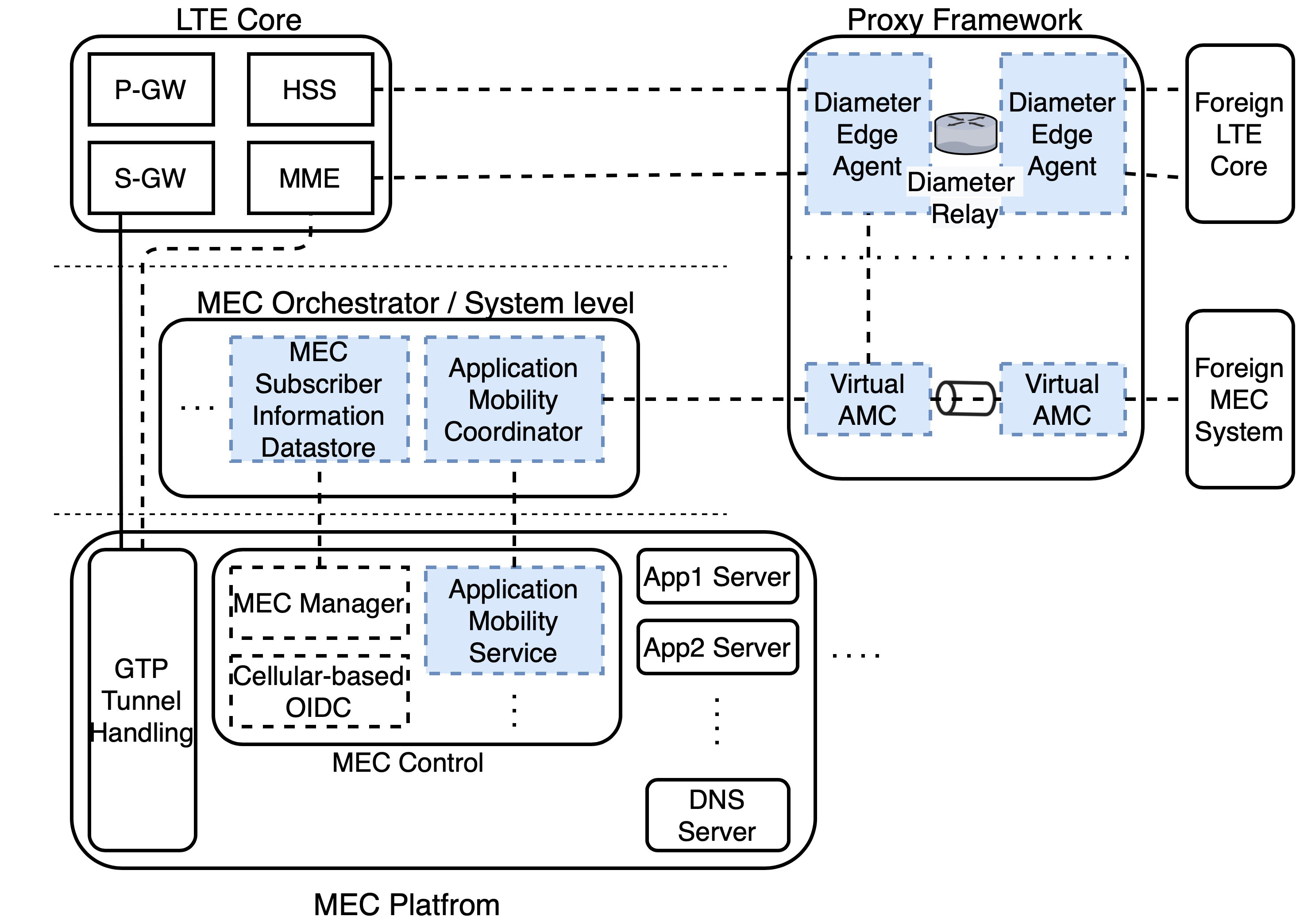}
\caption{Architecture design of a MEC system in the federation.}
\label{fig_solution}
\end{figure}

FS3A provides MEC-tier authentication and state transfer via a transparent proxy. The following components have been added or extended in the MEC system to accommodate FS3A: a system-wide datastore for each MEC system and mobility functional components, host-level AMC, and system level AMS. Fig. \ref{fig_solution} shows an MEC system with the relevant components for FS3A. The entities highlighted in blue are the ones we have added or extended in our proposed solution. The proxy system consists of distributed, interconnected proxy entities through which MEC systems and their underlying cellular networks connect to foreign networks via virtual counterparts so that the transparency is maintained. The MME and HSS communicate through the S6a interface which uses the Diameter protocol. In the MEC domain, virtual AMCs transfer messages related to user context from one network to another. To route the messages, the virtual AMCs have tables which store routing information for each of the networks.

\subsubsection{MEC Manager and MEO}
The MEC Manager and cellular-based OIDC module work according to MECSec design \cite{li2020transparent} with some additional functionalities for our federated design. The MEC Manager is responsible for activation and deactivation of MEC service for each UE. It fetches UE’s cellular credentials/attributes (IP, TEID, IMSI, subscription info etc.) from the associated LTE network or UE’s home network. The MEC subscriber datastore is a distributed database system placed inside each network and is used for storing and sharing identity and other UE information throughout the MEC system. They are incorporated along with the Mobile Edge Orchestrator (MEO) of the MEC system. The data are indexed using the most frequent IDs that are used. The data of UEs are kept in datastores near the UE’s location. These strategies allow for quicker access times.

\subsubsection{AMS and AMC}
The Application Mobility Service, or AMS, is the host level mobility function which exposes an interface to the MEC applications. The interfaces are provided by the applications for fetching the latest state information when requested by the MEC system. It is the task of the AMS to fetch state information from the applications where the UE was previously having a session and to provide the state information to the application where the UE has now moved . The AMC transfers the state information between different MEC platforms. When the state information is sent to a different network, it is passed through the proxy. The system-level mobility functions can be added to the MEC Orchestrator or can be placed into a separate component. For our proposed architecture, we have assumed that the MEC platforms have followed the MECSec design and have the AAA and access control components within them.

\subsection{Message flows}
The message flows for Federated 3rd-party authentication and state transfer in the FS3A procedure are elaborated as follows:

\begin{figure}[!t]
\centering
\includegraphics[width=3.3in]{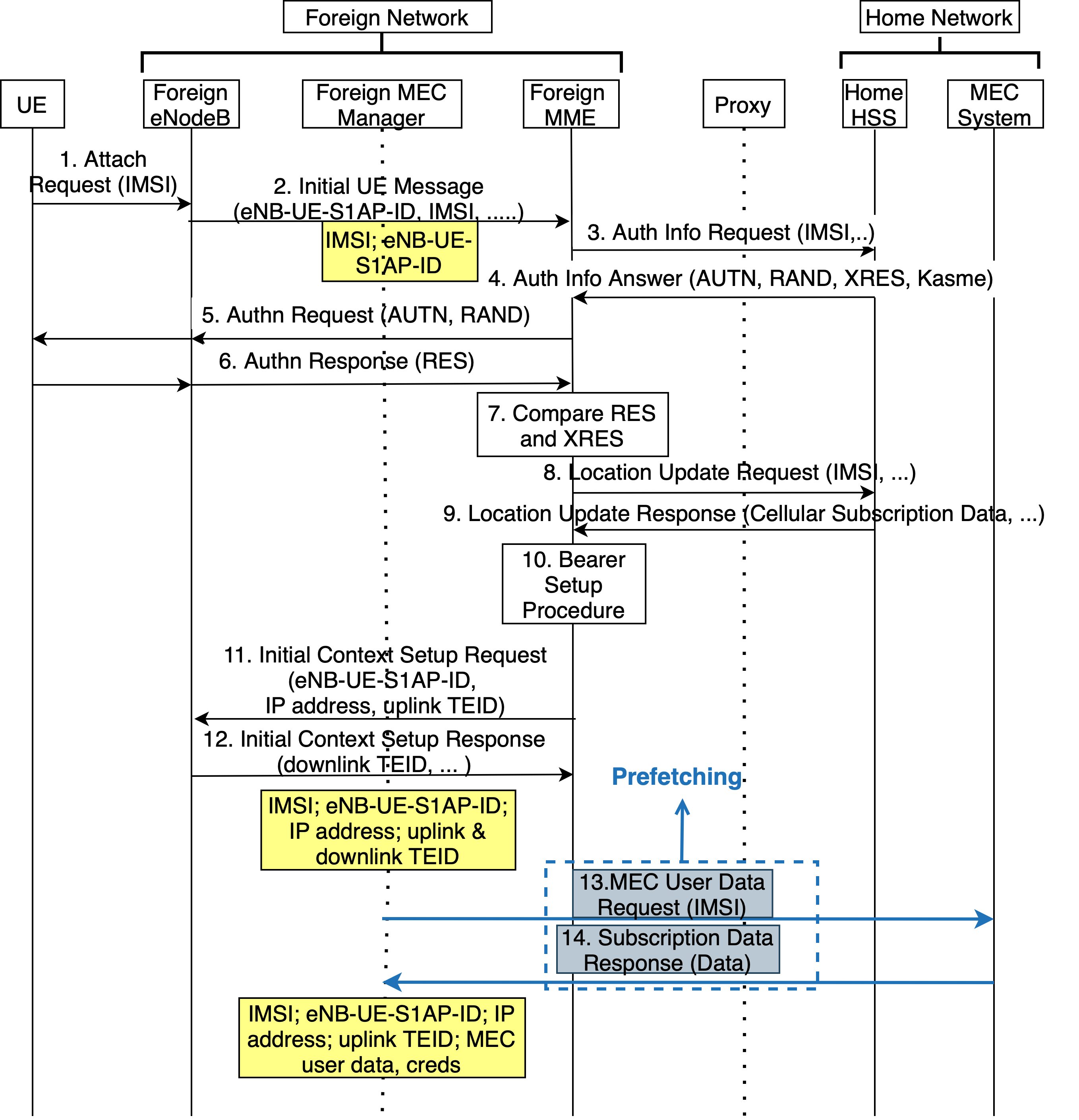}
\caption{Third-party Authentication Message Flow.}
\label{fig_auth_flow}
\end{figure}

\subsubsection{3rd-Party Authentication}
The 3rd party authentication for a foreign MEC application occurs in two stages which are UE identification stage and application authentication stage.

\emph{\textit{\textbf{UE identification.}}}
UE identification at a foreign network occurs when it attaches to that network for the first time. When the UE attaches to a foreign network during roaming, the authenticating MME and the home HSS are in two different networks and the authentication and location update messages travel across the networks through proxies. The S1 traffic between the MME and the eNB remains the same during the process, be it in a foreign network or in the home network. Fig. \ref{fig_auth_flow} shows that the MEC manager checks the S1 control plane traffic and checks the transferred cellular IDs during the initial attachment procedure to find out if any UE from another network has attached to the eNB where the MEC platform is deployed. When an UE attaches to network, the MEC manager gets the necessary ID of the UE from the S1 control plane packets and obtains the IMSI by inspecting the Initial UE Messages. The MEC manager maps the IMSI to a TEID address and IP address by looking at the Initial Context Setup Request. Hence, the MEC manager gathers all the necessary information about the UE in different stages as shown in Fig. \ref{fig_auth_flow}.

\begin{figure*}[!t]
\centering
\includegraphics[width=7in]{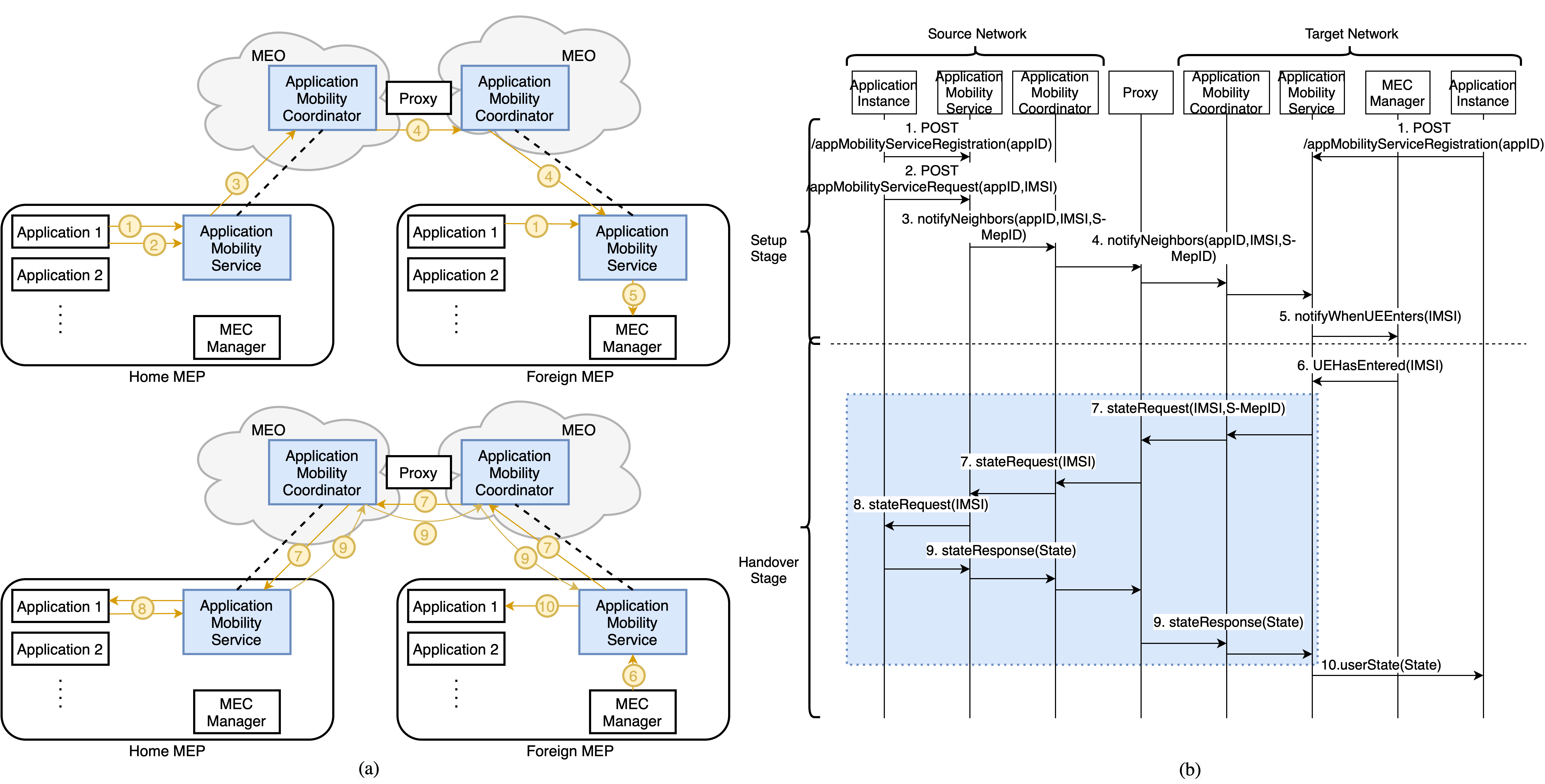}
\caption{(a) FS3A communication between entities: setup stage (top) and handover stage (bottom) (b) FS3A message flow.}
\label{fig_state_flow}
\end{figure*}

\emph{\textit{\textbf{Application authentication.}}} After the MEC system has identified the UE in the network, the cellular OIDC module can provide the identity of the UE to the requesting MEC application servers. The cellular OIDC modules then identify the UEs by mapping the source IP address of the authentication request packet sent by the UE with their IMSI. I If it is then found to be a valid user, the module provides the identity to the MEC server through the OIDC process. Then the MEC Application server provides the user with an authentication token. 

\subsubsection{State Transfer}
The state transfer portion of FS3A is also broken down into two stages, the setup stage and the handover stage. For application mobility we need the coordinated action of the AMS in the host level and the AMC in the system level. The AMS provides an interface for the MEC applications for accessing mobility services from the MEC system. Fig. \ref{fig_state_flow}(a) shows the FS3A communication between different entities for application mobility and Fig. \ref{fig_state_flow}(b) shows the message flow.

\emph{\textit{\textbf{Setup Stage.}}} If an application wants the UE state to be transferred when the UE changes platforms, the application informs the AMS beforehand (steps 1, 2). The source AMS then informs the system level AMC that an UE in its platform may transfer an application state to its neighbouring platforms (step 3). The AMC then informs about that application and the UE to the AMSs belonging to neighbouring MEC platforms of the source platform. The AMC may not be able to send this message to all neighbouring MEC platforms, especially if they are in other networks. So, through MEC Proxy, the AMC in the home network relays the message to the AMC in the foreign network which then completes the task (step 4). All the neighbouring AMSs are now informed about the MEC platform where the application state of that particular UE lies. If that UE enters any of those MEC platforms, they can then obtain its application state from the home MEC platform. The neighbouring AMSs also ask their MEC manager to inform them as soon as that particular UE attaches to the eNB of that MEC platform (step 5).

\emph{\textit{\textbf{Handover Stage.}}} When the UE enters any of the neighbouring MEC platforms, the MEC manager informs its AMS (step 6). The state transfer can take place as soon as the target MEC manager detects that the UE has connected to its network, even before the UE has set up a connection with the application. This is especially helpful when the state transfer is carried out between platforms in different networks. When an UE switches to a foreign network, a substantial amount of time is spent on authentication, network bearer setup, and connecting to the application. While all these processes are being carried out, the mobility functions can transfer the state to the right network. Thus, the extra time required for state transfer is negated. After being informed by the MEC manager about the arrival of the user in the network, the AMS fetches its application state from the source MEC platform. The request and response for state information is relayed through the AMC and proxies (steps 7-9). After the state information reaches the AMS, it provides the state information to its corresponding application (step 10).

\subsubsection{Optimizations} 
The delay induced by third-party authentication and state transfer can be reduced further with two optimization steps namely prefetching and token reuse. 

\emph{\textit{\textbf{Data Prefetching.}}} Some of the required subscription data can be fetched earlier from the source MEC in to the target MEC network, which we refer to as data prefetching. In our proposed solution, the target MEC requires the user subscription data from the source MEC. Although, this can be triggered after the completion of the EPS-AKA procedure, the foreign MEC manager can start prefetching this data as soon as it gets confirmation of the UE's authentication and has all the necessary information about the UE. The MEC manager maps the IMSI to a TEID address and IP address from the Initial Context Setup Request. So, this prefetching step can be started just after Initial Context Setup Request message in EPS-AKA as shown in blue box in Fig. 3.

In the state transfer solution, state data is fetched from the source MEC to the target MEC after the foreign MEC Manager informs foreign AMS about the UE's arrival. The state transfer starts after third-party authentication without data prefetching optimization but, just like the subscription data prefetching, state transfer can also be triggered by the foreign MEC manager as soon as it confirms the UE's successful authentication and has necessary UE's information. Thus, state transfer can be triggered just after the Initial Context Setup Request message in EPS-AKA. The state data prefetching is shown in the blue box in Fig. 4. Since the MEC system already parses S1 traffic for authentication, additional computational and memory loads for prefetching are negligible.

\emph{\textit{\textbf{Token reuse.}}} Cellular OIDC is used to provide long-term, secure, and reliable authentication and in cellular OIDC, after user's consent, Cellular OIDC module redirects the user to the MEC application with access/ID token. This token can be stored in the user's device so that when the user switches MEC network, the token could be submitted to the foreign MEC network, thus skipping the Cellular OIDC procedure. Reusing tokens allows us to skip over the cellular OIDC procedure, and thus reduces the delay of service resumption.

\begin{figure*}[!t]
\centering
\includegraphics[width=6in]{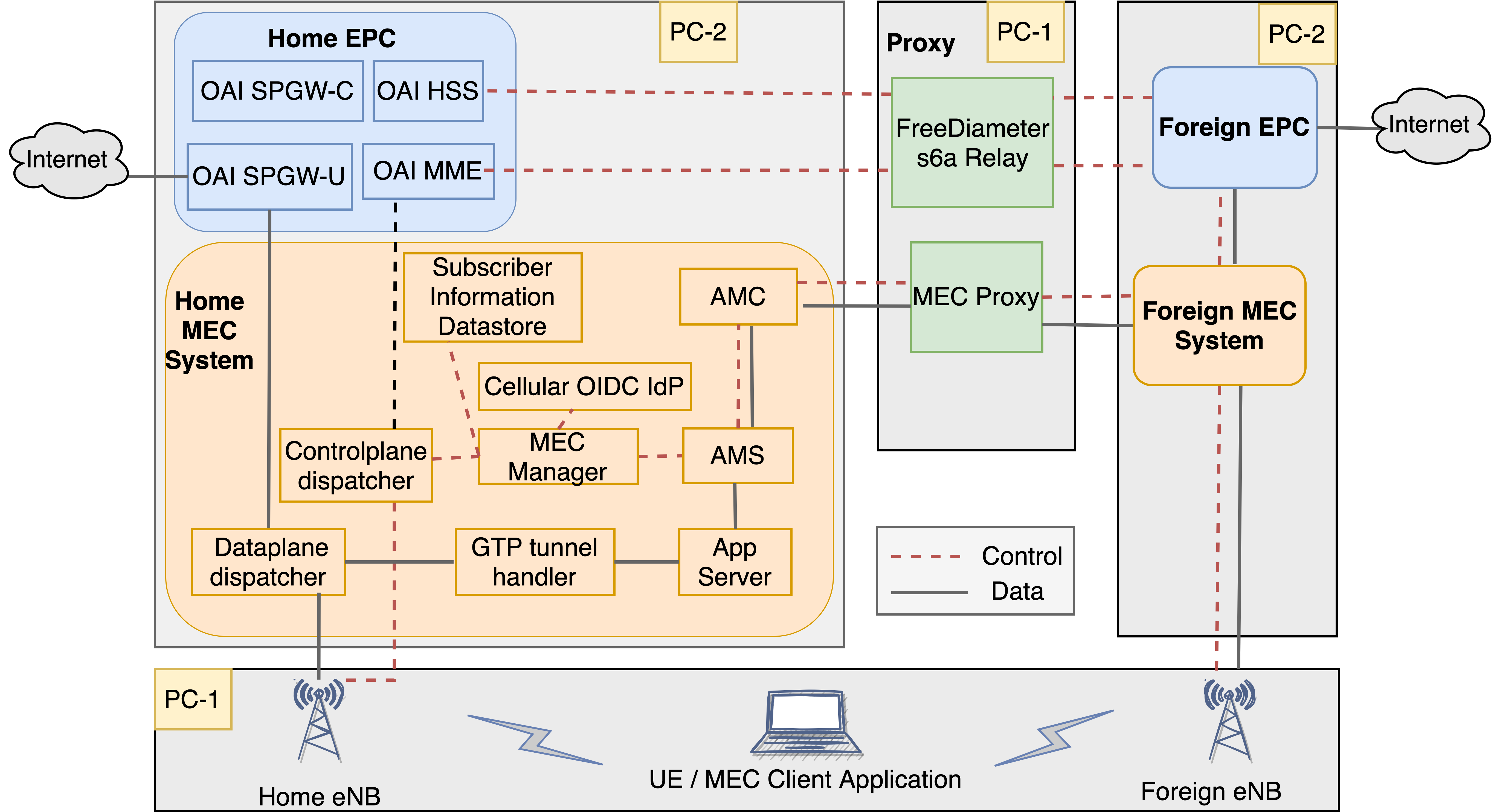}
\caption{Experimental Testbed.}
\label{fig_five}
\end{figure*}

\subsubsection{Security Analysis}
We assumed that the MEC platforms themselves follow the MECSec design so that they have access control systems that protect against malicious traffic. The UE is authenticated in a foreign MEC system only after it has first been authenticated by the foreign cellular network . Since the MEC system uses the source IP address of a packet to identify the user that has sent it, a malicious user may try to spoof a valid IP address to get access into the network. This is also a threat to the underlying cellular network. The cellular network attempts to mitigate this by changing the IP address of the user from time to time, making it harder for attackers to track  addresses. The MEC system keeps track of these changes by inspecting the S1 control plane messages. Thus, it is always informed about legitimate users.

The application authentication tokens may also pose a security threat as malicious users may try to generate fake tokens or tamper with tokens, or even reuse valid tokens meant for other users. Tokens are themselves made secure since they are signed and encrypted by keys only available to the applications. And, the lifetime of the tokens is also very short. Again, a malicious user managing only the token cannot gain entry into the MEC system, as the MEC system will inspect the IP address and determine that it is not a legitimate user. The proxy is a separate network which cannot be accessed by outside users; it is only accessed by trusted networks over secure connections. So, malicious users cannot hack into the proxy, and overall, FS3A is resilient against malicious attacks.

\section{Implementation}
In this section we  cover the design of our experimental prototype, implemented modules, and the testbed used for experimentation.

\subsection{Prototype Architecture}
We test our proposed solution on an OpenAirInterface (OAI) based 4G-LTE network. The solution is equally feasible for the 5G cellular network as it does not propose any changes to the underlying cellular network architecture. We set up two networks for the prototype, each consisting of an LTE cellular network and an MEC system integrated with it. To run the experiment, we set up an MEC application server in each network. The MEC applications authenticate their users through cellular OIDC, and behave the same as any OIDC Client. The cellular OIDC Identity Provider provides IMSI information to the MEC application as the user ID. The proxy consists of a FreeDiameter s6a relay and a custom relay for MEC information transfer.

\subsection{Modules}
OpenAirInterface \cite{openairinterface} implementations provide the LTE cellular network platform. We  used a ueltesimulator \cite{gitlab} to simulate the UE. Roaming support was added to the LTE networks so that S6a traffic of roaming users were redirected to their home networks. The dataplane dispatcher routes the MEC data towards the MEC servers and the GTP tunnelling module decapsulates the data packets before they are sent to the MEC servers. libgtpnl library \cite{osmocom} was used to handle GTP tunnels. The control plane dispatcher, implemented in C, sits between the S1AP interface between the eNB and the MME and acts as a proxy. It sends a copy of the packets to the MEC Manager.

\begin{table*}[!t]
\renewcommand{\arraystretch}{1.3}
\caption{Different scenarios created via multiple combinations of Auth server location, state prefetching and token reuse.}
\label{table_three}
\centering
\begin{tabular}{|c||c||c||c|}
\hline
Scenarios & Authentication Server Location & Timing for Fetching subscription data & Token Reuse or Re-authentication\\
\hline
CUA & Cloud(C) & During User Auth (U) & Re-authentication (A)\\
\hline
CUT & Cloud(C) & During User Auth (U) & Token Reuse (T)\\
\hline
CPA & Cloud(C) & Prefetching (P) & Re-authentication (A)\\
\hline
CPT & Cloud(C) & Prefetching (P) & Token Reuse (T)\\
\hline
MUA & MEC(M) & During User Auth (U) & Re-authentication (A)\\
\hline
MUT & MEC(M) & During User Auth (U) & Token Reuse (T)\\
\hline
MPA & MEC(M) & Prefetching (P) & Re-authentication (A)\\
\hline
MPT & MEC(M) & Prefetching (P) & Token Reuse (T)\\
\hline
\end{tabular}
\end{table*}

The MEC Manager was implemented with python via using the pycrate \cite{github} library, and it parses the messages and sends the information relevant for MEC authentication to the cellular OIDC module, which in turn accumulates them in a Dictionary. The cellular OIDC module, AMS, AMC and the MEC server application  were implemented using Node.js. The front end application and back end MEC server applications are simple React and Express Node.js applications respectively. The MEC relay component of the proxy was implemented using Node.js, and the information of different networks was stored and managed by this component. They have persistent socket connections in between them for faster data transfers. The detailed architecture with all the modules are shown in Fig. \ref{fig_five}.

\subsection{Testbed}
Our testbed was  set up on 2 PCs, both with same hardware configuration and Linux Ubuntu Operating System. The UE and eNB of both cellular networks was set up in PC-1, while PC-2 was set up with the EPC and MEC components. Proxy components were set up in PC-1 to make communication between two EPC more realistic. Docker was used to containerize the components and docker networks were used to create virtual networks in each PC. Both the PCs were connected to the same LAN through a router/switch. We also deployed the authentication server which was an OIDC IdP provider written in Node.js. The cloud components were hosted in Google Cloud Platform (GCP) in order to test the latency difference between the scenario where the authentication server was deployed in the cloud and the scenario where the authentication server was deployed in the MEC. The location of the GCP server was set at asia-southeast-1 (Singapore) which was the closest available server to the rest of the testbed in Bangladesh. The bandwidth of the network between the MEC setup and the cloud was 20Mbps.

\section{Results and Evaluation}
After implementation, we evaluated our proposed FS3A design to  ascertain the time taken for the UE authentication in different scenarios.  We then examined the state transfer latency for different state sizes and  divided the UE authentication and state transfer latency into multiple stages to determine the effectiveness of the proposed token reuse and data prefetching optimizations. Finally, we investigated the service interruption latency for the different scenarios to ascertain the overall effectiveness of FS3A. 

\subsection{UE Authentication Latency}
For the first set of results, we investigated the UE authentication latency by dividing it into three steps: authentication with the foreign MEC, access control, and registration. We also considered multiple scenarios to calculate the UE authentication latency in order to determine the efficiency of FS3A. The authentication time was calculated for 8 possible scenarios, as shown in Table \ref{table_three}. We designed these scenarios by placing the authentication server in the cloud or the MEC, by fetching the subscription data during UE authentication or prefetching, and by carrying out the UE re-authentication at the target MEC or reusing the authentication token provided to the UE by the application server in the home MEC.

\begin{figure}[!t]
\centering
\includegraphics[width=3.3in]{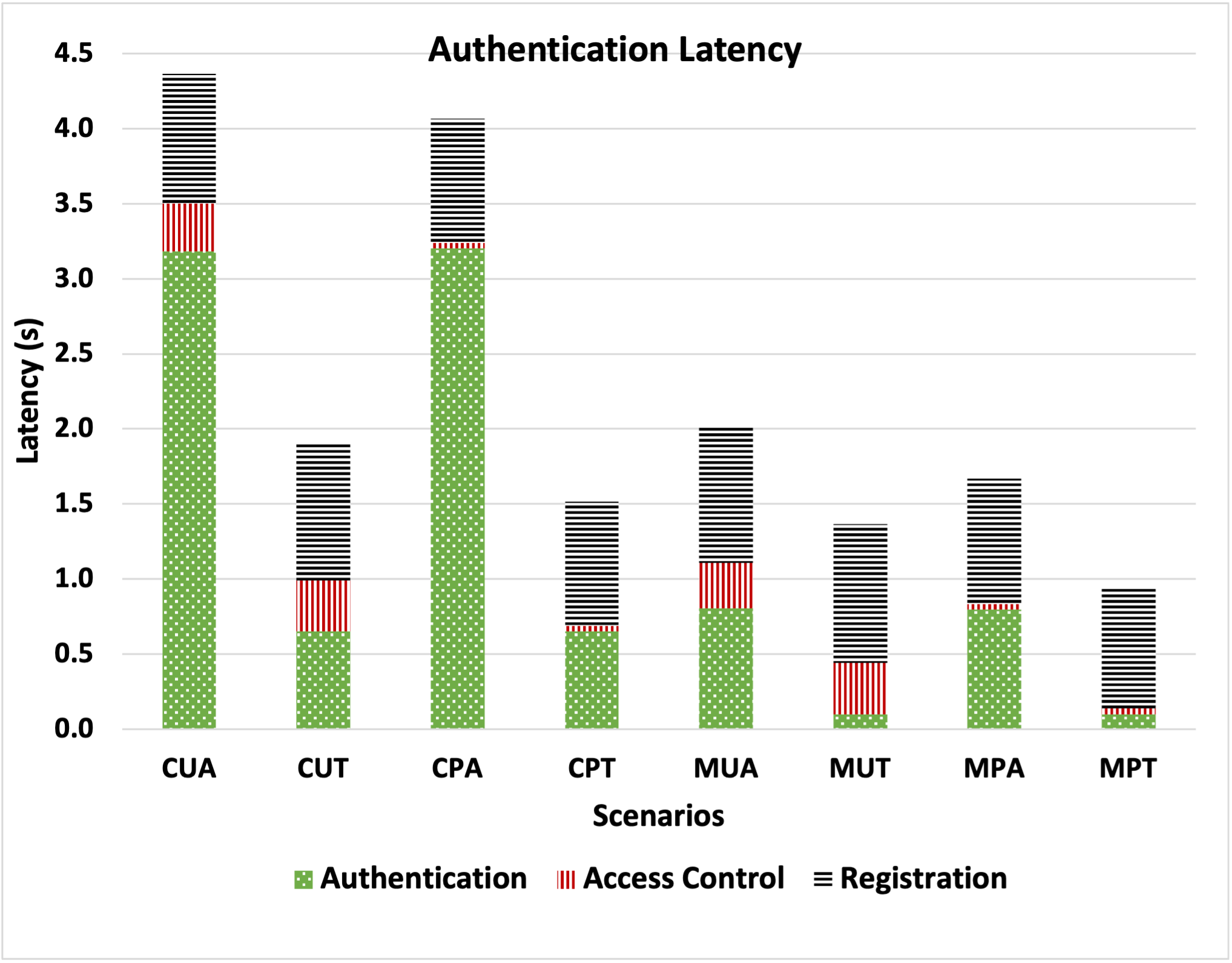}
\caption{UE Authentication Latency.}
\label{fig_six}
\end{figure}

We calculated the UE authentication latency for all 8 scenarios, as can be seen in Fig. \ref{fig_six}. It was evident that the MPT scenario (FS3A) took the least amount of time as it used the subscription data prefetching and reused the authentication tokens while the authentication server was in the MEC. The results also show that placing  the authentication server in the MEC reduced the latency by 28\textendash58\% compared to placing  the authentication server in the cloud. It was also seen that the token reuse and the subscription data prefetching reduced the authentication latency by 53\textendash65\%,  compared to the complete re-authentication and subscription data fetching during UE authentication. This clearly shows the importance of the proposed token reuse and data prefetching optimizations. 

\subsection{State Transfer Latency}
For the second set of results, we calculated the state transfer latency to see how much time was taken to fetch the application state from the UE’s home network in order to resume the application in the target network. We measured this latency for the states of different sizes in three different cases. In the first case, the state was downloaded from the cloud, which might be required if there is no FS3A in place to transfer the state from one MEC to another MEC. In the second case, the state was transferred via MEC proxy without using the state data prefetching, while in the third case, state was transferred via MEC proxy along with the state data prefetching.

\begin{figure}[!t]
\centering
\includegraphics[width=3.3in]{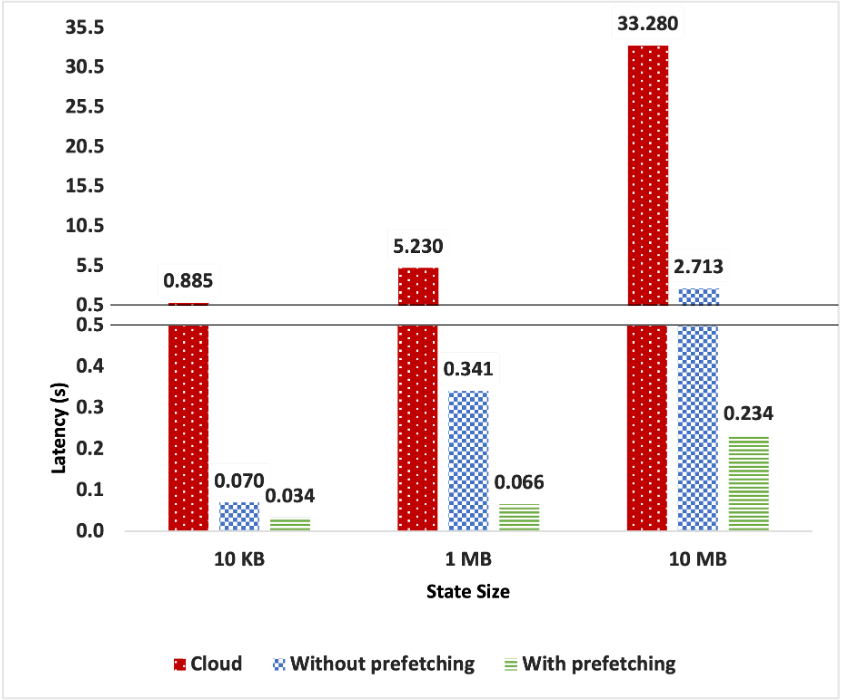}
\caption{State Transfer Latency.}
\label{fig_seven}
\end{figure}

Fig. \ref{fig_seven} shows the state transfer latency comparison of the above mentioned three cases for the state sizes of 10 KB, 1 MB, and 10 MB. It is clear from  Fig. \ref{fig_seven} that, as the states size increased, the latency for the state transfer via the cloud server during handover increased considerably. Therefore, we suggest that Mobile Network Operators avoid using the cloud for state transfer as it leads to poor user experience. Our proposed FS3A (without state prefetching) took much less time  compared to  state transfer via the cloud as it transferred the state via the MEC proxy. It can also be seen that, by using state prefetching, we further reduced the state transfer latency by 51.4\%, 80.6\% and 91.3\% for state sizes of 10 KB, 1 MB, and 10 MB, respectively, compared to state transfer without prefetching.

\subsection{Latency Breakdown with vs. without Optimizations} 
\begin{figure}[!t]
\centering
\includegraphics[width=3.3in,height=2.7in]{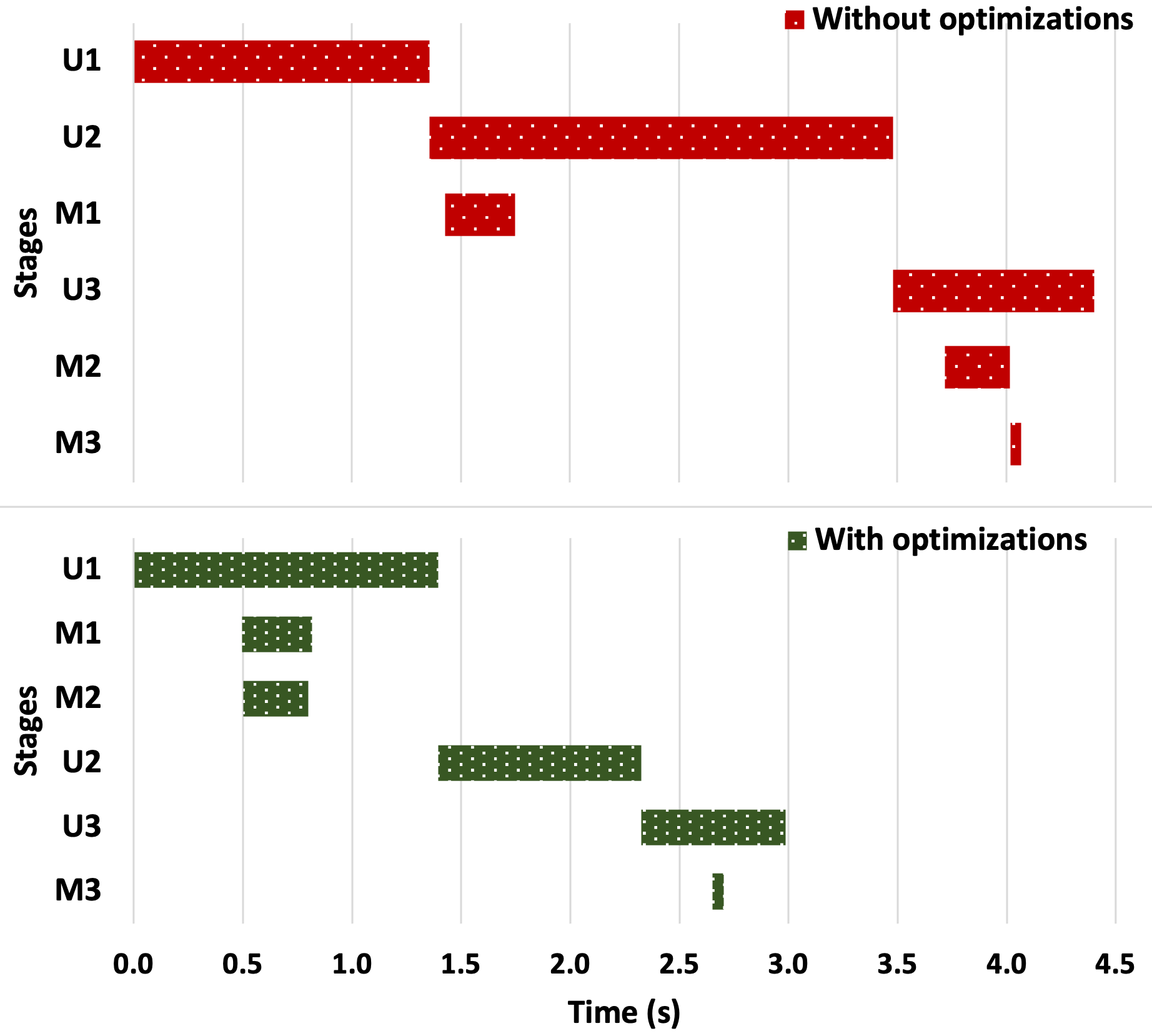}
\caption{Latency Breakdown With vs. Without Optimizations.}
\label{fig_eight}
\end{figure}

For the third set of results, we analyzed the overall application resumption latency for the UE with and without our proposed token reuse and data prefetching optimizations. The key stages involved in the MEC application resumption are listed in Table \ref{table_four}. The UE attached to the MEC in the U1 stage, became authenticated in the U2 stage, and application was resumed for the UE in the U3 stage. The U2 stage required the MEC in the foreign network to fetch the subscription data from the UE’s home network (M1). Similarly, the U3 stage required the MEC to fetch the state data from the UE's home network (M2). M3 stage was not involved in the MEC application resumption time and was needed for the UE to move to another MEC later. Fig. \ref{fig_eight} shows the time taken by each stage for the two scenarios; without and with our proposed token reuse and data prefetching optimizations.

\begin{table}[!t]
\renewcommand{\arraystretch}{1.3}
\caption{Stages in MEC application resumption.}
\label{table_four}
\centering
\begin{tabular}{|c||c||c||c|}
\hline
Name & Stage involving UE & Name & Satge involving MEC\\
\hline
U1 & UE Attach & M1 & Fetching Subscription Data\\
\hline
U2 & UE Auth & M2 & Fetching State Data\\
\hline
U3 & Application & M3 & Notifying Neighbour MECs\\
 & Resumption & & \\
\hline
\end{tabular}
\end{table}

In both the scenarios, the authentication server was located in the MEC and the state size was 1 MB. Without optimizations, the M1 stage took place during the U2 stage while the M2 and the M3 took place during the U3 stage. As Fig. \ref{fig_eight} shows, the M1 and the M2 stages induced the additional latency of 321 ms and 297 ms, respectively. We started the M1 and the M2 stages earlier during the U1 stage to reduce the time taken by the U2 and  U3 stages, as seen in the lower half of  Fig. \ref{fig_eight}. It can also be seen that the M1 and  M2 stages were completed before they were needed by the U2 and the U3 stages. The reuse of the authentication token of the app server further reduced the latency of the U2 stage. The rest of the time taken by the U2 and the U3 stages was mainly because of the communication between UE and application server  which  could not be further reduced. Therefore, we suggest that mobile network operators  deploy the proposed token reuse and  data prefetching optimizations as they reduce the latency of the U2 and the U3 stages by 56.3\% and 28.3\%, respectively.

\subsection{Service Interruption Latency}
When the UE detached from one MEC and attached to the MEC in another MNO, the service was interrupted until the application was for the UE resumed. The service interruption latency was divided into four stages: UE attach, UE authentication, MEC to MEC communication for state transfer, and MEC to UE communication. We calculated this service interruption latency for three different scenarios described in Table \ref{table_five}. The first scenario considered the authentication server in cloud and transferred the state via the cloud without data prefetching and token reuse optimizations. The second scenario considered the authentication server in MEC and transferred the state via MEC proxy without data prefetching and token reuse optimizations. The third scenario considered the authentication server in MEC and transferred the state via MEC proxy with data prefetching and token reuse optimizations (i.e. our proposed FS3A).

\begin{figure}[!t]
\centering
\includegraphics[width=3.3in]{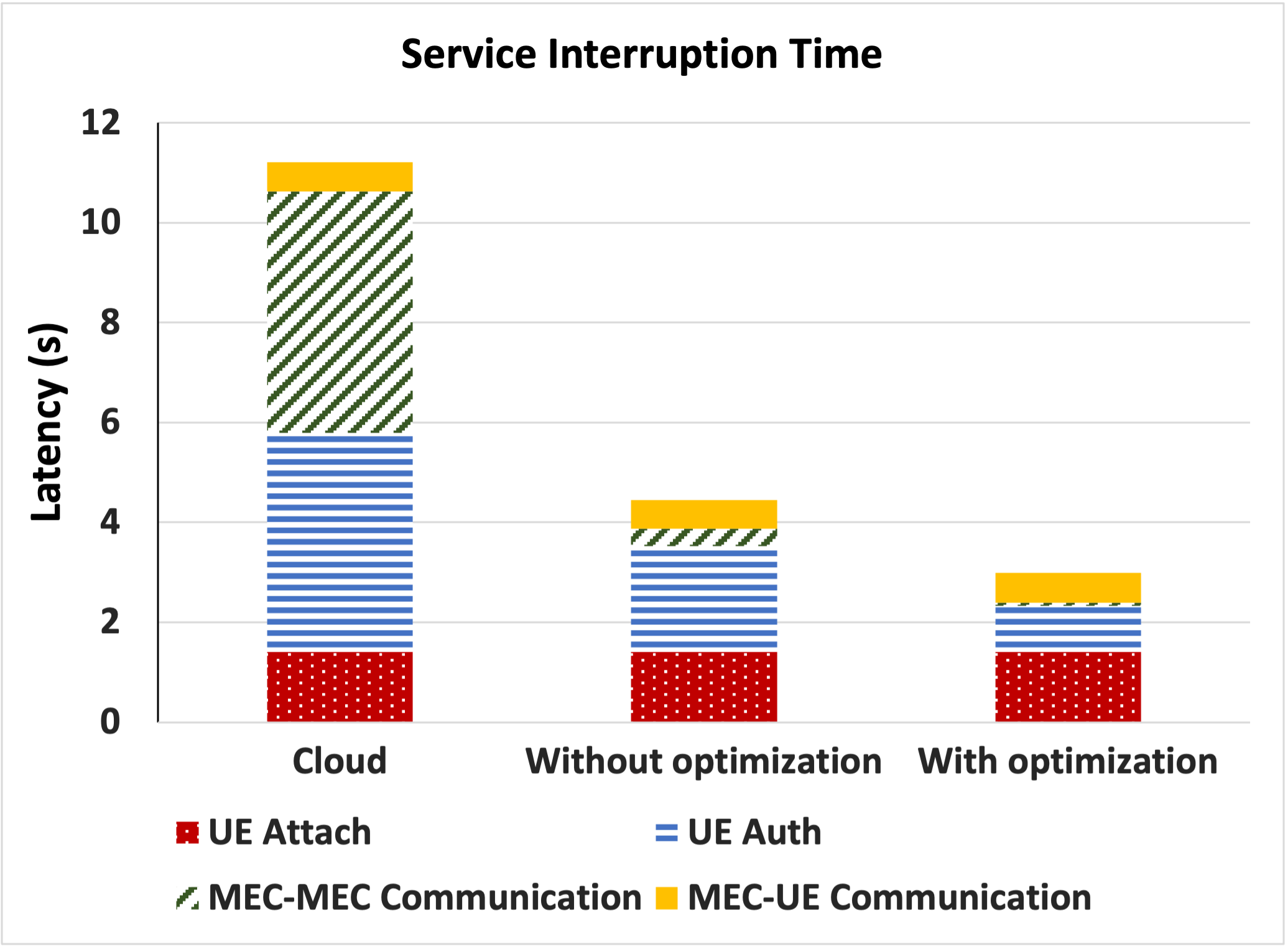}
\caption{Service Interruption Latency Comparison.}
\label{fig_nine}
\end{figure}

The comparison of service interruption latency for these three scenarios is shown in Fig. \ref{fig_nine}, which shows that, with our proposed architecture and token reuse and data prefetching optimizations (scenario 3), service interruption latency was 2.96 seconds, that is 73.6\% and 33.1\% less  compared to the scenarios 1 and 2 respectively as a result of the MEC proxy, token reuse, and data prefetching. It should be noted that the UE attach and the MEC to UE communication stages took almost the same amount of time in all three scenarios as they mostly depended upon  propagation delay and could not be reduced. It can be seen that we considerably reduced the UE authentication time and MEC to MEC communication time by token reuse and data prefetching optimizations. If we ignore the time taken by the UE attach and the MEC to UE communication stages, the latency for scenario 3 can be further be reduced by 59.7\%  compared to the scenario 2, which is a considerable decrease in service interruption latency.

\begin{table}[!t]
\renewcommand{\arraystretch}{1.3}
\caption{Scenarios for measuring service interruption time.}
\label{table_five}
\centering
\begin{tabular}{|c||c||c||c|}
\hline
Scenario & Server & State Transfer & Data Prefetching\\
 & Location & via & \& Token Reuse\\
\hline
1 & Cloud & Cloud & No\\
\hline
2 & MEC & MEC Proxy & No\\
\hline
3 & MEC & MEC Proxy & Yes\\
\hline
\end{tabular}
\end{table}

\section{Conclusions and Future Work}
MEC is one of the most imminent technologies in 4G/5G networks as it brings computational services closer to  end users. MECs are deployed by various mobile network operators and, in future, mobile users will have to face  authentication and application mobility issues when they move across from one MNO to another MNO. It would be a tedious task to buy subscriptions from multiple MNOs in order to benefit from a continuous MEC experience. In order to address these issues, we propose FS3A mechanism for third-party authentication and low-latency state transfers among MECs and across different MNOs. FS3A makes use of a transparent proxy to provide seamless and fast 3rd-party authentication and application mobility while achieving low latency via authentication token reuse and subscription, and state data prefetching. The results show that:
\begin{itemize}
    \item \emph{\textit{\textbf{Reusing  authentication tokens reduces  authentication latency.}}} FS3A saves 2543 ms and 701 ms on average by reusing the tokens from authentication servers in cloud and MEC respectively.
    \item \emph{\textit{\textbf{Subscription data prefetching reduces the access control latency.}}} FS3A reduces the access control latency by 88.6\% by prefetching the subscription data.
    \item \emph{\textit{\textbf{State data prefetching becomes crucial as the state size increases.}}} State transfer latency is reduced by 51.4\textendash91.3\% for state of size of 10KB\textendash10MB via state prefetching.
    \item \emph{\textit{\textbf{Token reuse and prefetching play an important role in latency reduction.}}} Token reuse and prefetching optimizations resume the application by taking 33\% less time  compared to no token reuse and prefetching.
\end{itemize}

This work addressed the authentication and application handover problem between the MECs that belong to multiple cellular service providers and proposed a horizontal federation among MECs. In future, we will extend this work to form a federation among different service providers such as cloud, edge, and fog in order to create vertical and hybrid federations. This work can also be extended in another direction by considering  other federation issues, apart from authentication and application mobility, such as traffic offloading, load balancing and capacity sharing among multiple MNOs.

\begin{IEEEbiography}[{\includegraphics[width=1in,height=1.25in,clip,keepaspectratio]{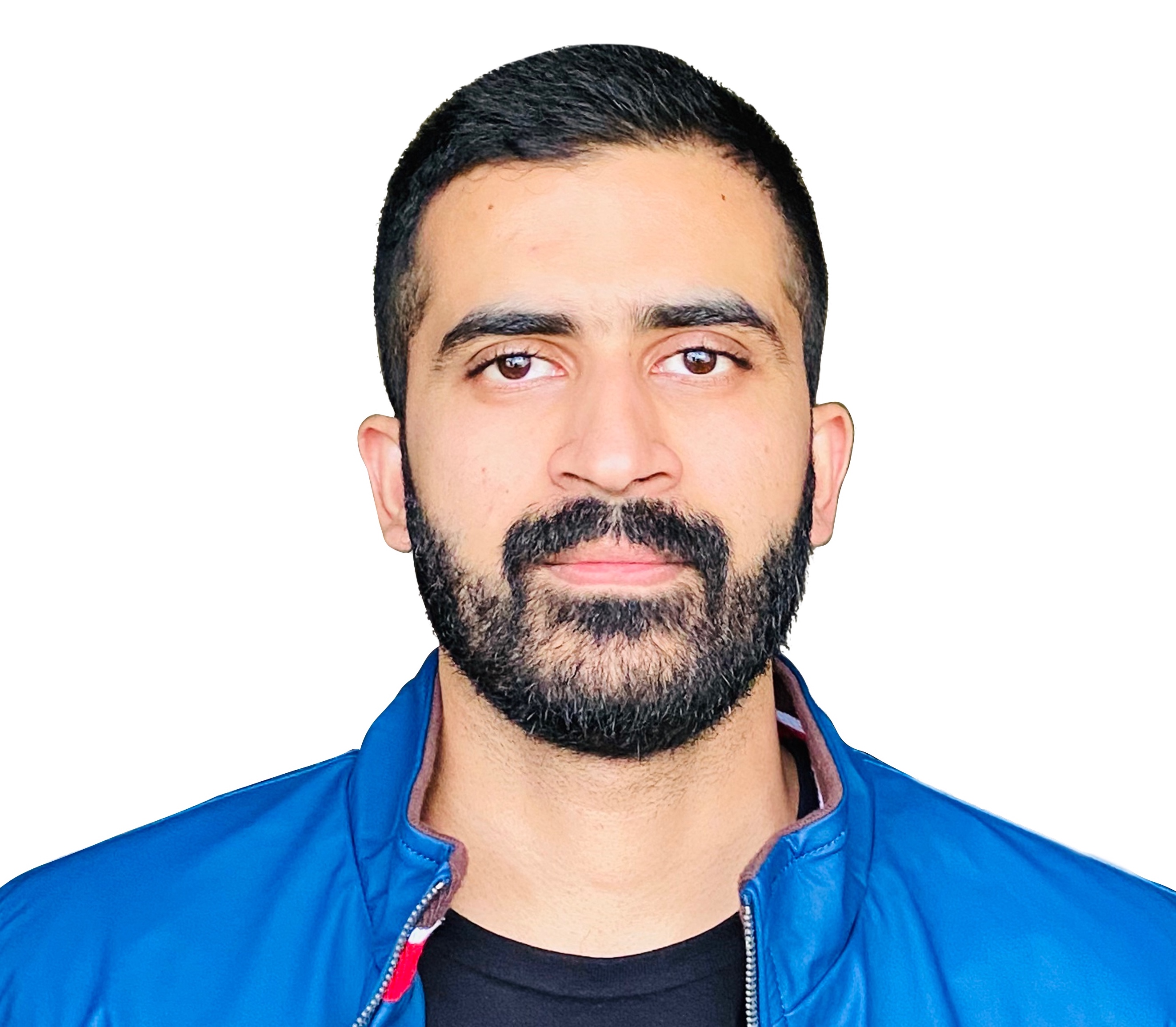}}]{Asad Ali} received his BS degree in electrical engineering from the University of Engineering and Technology, Taxila in 2012. In 2015, he received his Master degree in Electrical Engineering from National University of Science \& Technology (NUST), Pakistan. At the moment, he is a Ph.D. researcher in the Electrical Engineering and Computer Sciences department of the National Yang Ming Chiao Tung University (NYCU), Taiwan. His research interests include network security, network protocols, wireless communications, artificial intelligence wireless, network design, and optimization.
\end{IEEEbiography}

\begin{IEEEbiography}[{\includegraphics[width=1in,height=1.25in,clip,keepaspectratio]{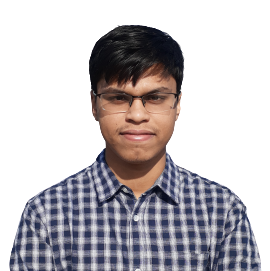}}]{Samin Rahman Khan} is a final year undergraduate student completing his B.S.Engg in Computer Science and Engineering from Bangladesh University of Engineering and Technology (BUET), Bangladesh. He has stepped into research, through working on Multi-Access Edge Computing and Mobile Networking technologies. He is interested in the fields of Computer Networking, Security, and Applied Machine Learning.
\end{IEEEbiography}

\begin{IEEEbiography}[{\includegraphics[width=1in,height=1.25in,clip,keepaspectratio]{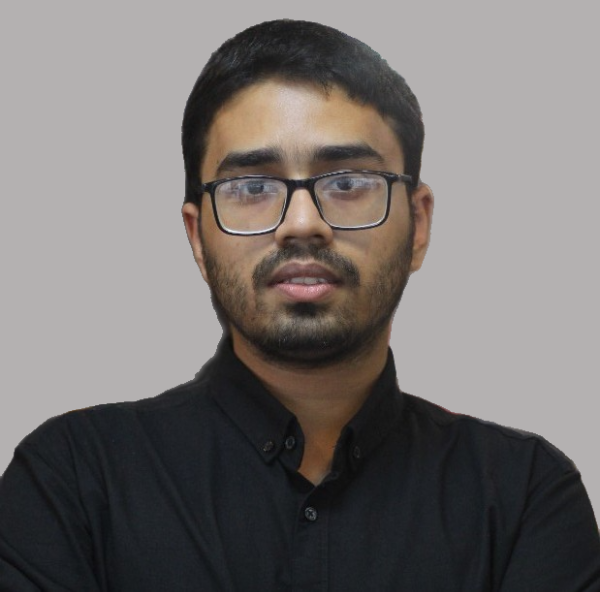}}]{Sadman Sakib} is currently pursuing BS degree in Computer Science and Engineering at Bangladesh University of Engineering and Technology (BUET), Bangladesh. He is expected to graduate in 2022. His research interests include edge computing, IoT security, computer networks, and applied machine learning.
\end{IEEEbiography}

\begin{IEEEbiography}[{\includegraphics[width=1in,height=1.25in,clip,keepaspectratio]{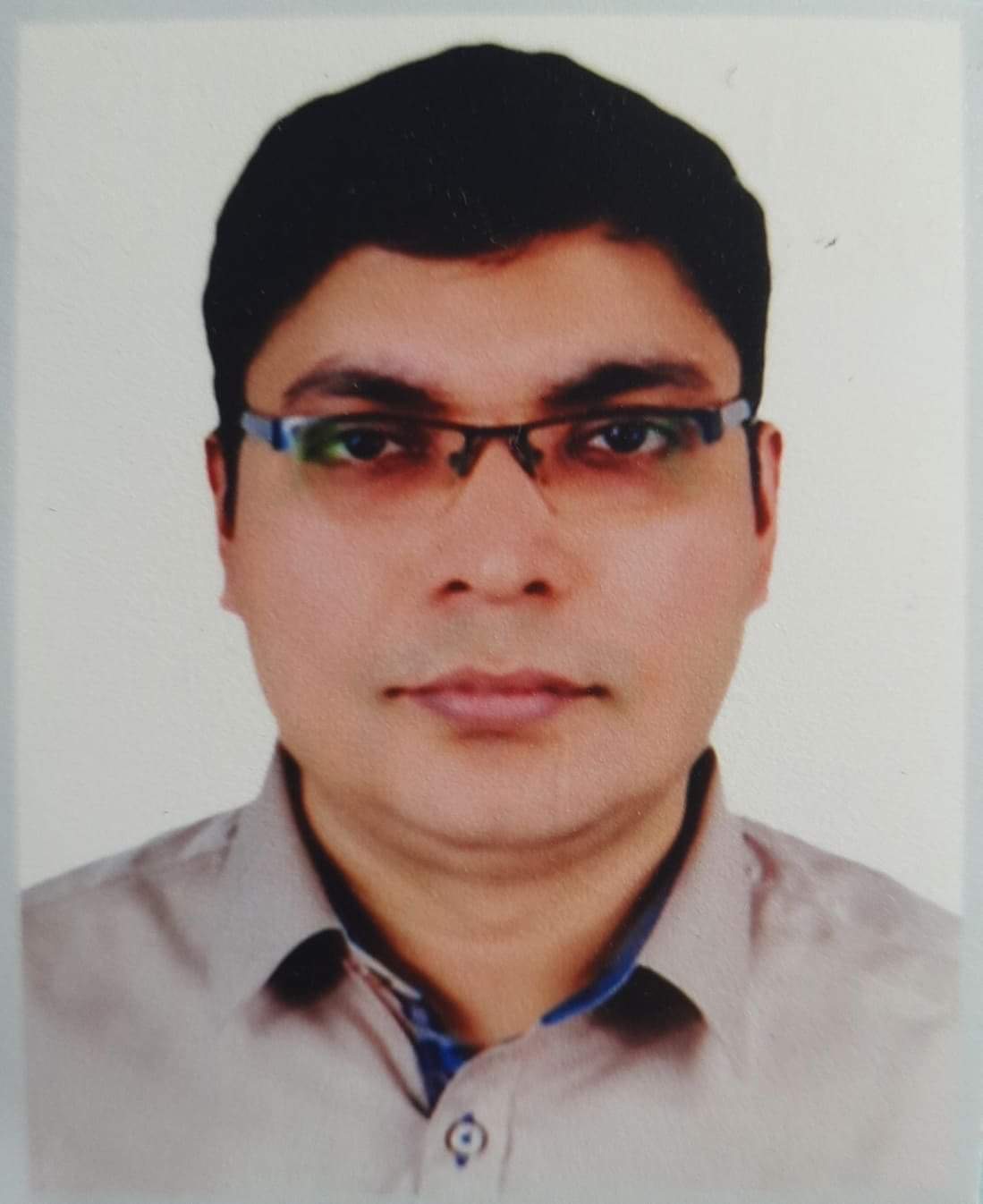}}]{Md. Shohrab Hossain} is a Professor of Computer Science and Engineering at Bangladesh University of Engineering and Technology (BUET), Bangladesh. He received his Ph.D. degree in Computer Science from the University of Oklahoma, USA in 2012. His research interests include Mobile malware detections, cyber security, Software defined networking (SDN), security of mobile and ad hoc networks, and Internet of Things. He has published more than 75 technical research papers in leading journals and conferences. He has been serving as the TPC member of IEEE GLOBECOM, IEEE ICC, IEEE VTC, Wireless Personal Communication, (Springer), Journal of Network and Computer Applications (Elsevier), IEEE Wireless Communications.
\end{IEEEbiography}

\begin{IEEEbiography}[{\includegraphics[width=1in,height=1.25in,clip,keepaspectratio]{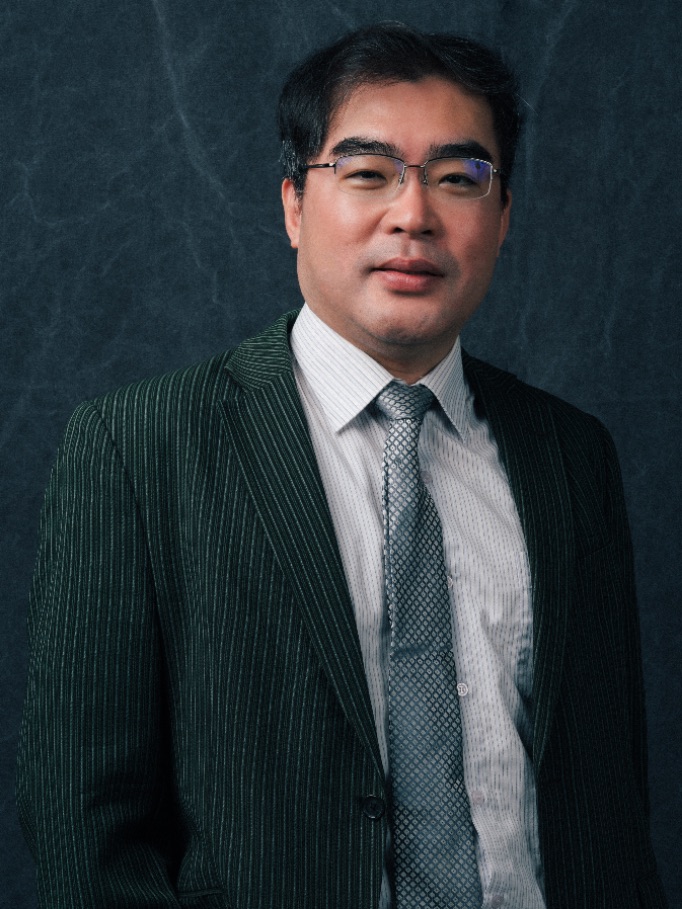}}]{Ying-Dar Lin} is a Chair Professor of computer science at National Yang Ming Chiao Tung University (NYCU), Taiwan. He received his Ph.D. in computer science from the University of California at Los Angeles (UCLA) in 1993. Since 2002, he has been the founder and director of Network Benchmarking Lab. He has served or is serving on the editorial boards of several IEEE journals and magazines, and was the Editor-in-Chief of IEEE Communications Surveys and Tutorials (COMST) during 2017-2020. He published a textbook, Computer Networks: An Open Source Approach His research interests include network security, wireless communications, and network softwarization.
\end{IEEEbiography}

\EOD

\end{document}